\documentclass[conference]{IEEEtran}

\usepackage{cite}
\usepackage{amsmath,amssymb,amsfonts}
\usepackage{algorithmic}
\usepackage{algorithm}
\usepackage{graphicx}
\usepackage{textcomp}
\usepackage{xcolor}
\usepackage{url}

\usepackage{times}
\usepackage{amsmath}
\usepackage{amssymb}
\usepackage{booktabs}
\usepackage{graphicx}
\usepackage[hidelinks]{hyperref}
\usepackage{url}
\usepackage{xcolor}
\usepackage{microtype}
\usepackage{authblk}
\usepackage[numbers,sort&compress]{natbib}
\usepackage{titlesec}
\usepackage{abstract}
\usepackage{placeins}
\usepackage[font=footnotesize,labelfont=bf,skip=4pt]{caption}
\usepackage{subcaption}
\usepackage{tikz}
\usepackage{comment}
\titleformat{\section}{\normalfont\large\bfseries}{\thesection.}{0.5em}{}
\titleformat{\subsection}{\normalfont\normalsize\bfseries}{\thesubsection.}{0.5em}{}
\titleformat{\paragraph}[runin]{\normalfont\normalsize\bfseries}{}{0em}{}[.]
\titlespacing{\section}{0pt}{6pt}{3pt}
\titlespacing{\subsection}{0pt}{4pt}{2pt}

\usetikzlibrary{arrows.meta, positioning}

\title{Comparative Analysis of State-of-the-Art Foundation Models for Sleep Analysis Under Channel Reduction }

\author[1]{Hassan Mehdi}
\author[2]{Riku Kl\'en}
\author[1]{Ayse Kosal Bulbul}
\author[3]{Suzanne Timmons}
\author[1,4]{Abdulhamit Subasi}
\author[5]{Wei Chen}
\author[6]{Zou Zhu}
\author[1,2,5,6]{\protect\\ Muhammad Irfan}

\affil[1]{\small Institute of Biomedicine, University of Turku, Finland}
\affil[2]{\small Turku PET Centre, University of Turku and Turku University Hospital, Finland}
\affil[3]{\small University College Cork, Ireland}
\affil[4]{\small AI4HEALTH Lab. University at Albany, SUNY,  Albany, NY, USA}
\affil[5]{\small Fudan University, Shanghai, China}
\affil[6]{\small School of Biomedical Engineering, University of Sydney, Australia}

\date{\small 27 July, 2026}

\begin{document}

\maketitle

% Dataset reference block — added for readers following this link

 \begin{comment}
\vskip -6pt
\noindent\colorbox{gray!12}{\parbox{0.97\linewidth}{%
\small\textbf{[59] Dataset References}\\[2pt]
\textbf{NSRR:}
MESA~\url{https://sleepdata.org/datasets/mesa},
SHHS~\url{https://sleepdata.org/datasets/shhs},
MrOS~\url{https://sleepdata.org/datasets/mros}\\[2pt]
\textbf{PhysioNet:}
Sleep-EDF~\url{https://physionet.org/content/sleep-edfx/1.0.0/},
DREAMT~\url{https://physionet.org/content/dreamt/2.2.0/},
HMC~\url{https://physionet.org/content/hmc-sleep-staging/1.1/}\\[2pt]
\textbf{ISRUC-Sleep:}~\url{https://sleeptight.isr.uc.pt/}
}}
\vskip 6pt
\end{comment}

\begin{abstract}

Automatic sleep staging from polysomnography (PSG) is a well-studied task, but PSG itself is expensive, clinic-based, and burdensome to manually score, which limits its use for long-term or at-home monitoring. Most existing sleep-staging foundation models are evaluated using the full PSG montage. We instead ask how much of that montage is actually necessary. We evaluate six sleep staging models on the Multi-Ethnic Study of Atherosclerosis (MESA) PSG dataset across three signal conditions: electroencephalography (EEG), electrocardiography (ECG), and their combination (EEG+ECG). This is motivated by edge-cloud deployment, where EEG requires a clinic-grade scalp electrode, whereas ECG is already captured by consumer wearables. We test state-of-the-art foundation models such as SleepFM with an encoder trained from scratch on MESA, alongside BIOT, MOMENT, LaBraM, a base-scale Vision Transformer (ViT-B) reimplementation of SensorLM trained from scratch, and YASA, spanning EEG-pretrained, general-time-series, from-scratch, and classical non-learned approaches. No model architecture is modified from its original form; SensorLM's encoder is reimplemented only in PyTorch. For EEG-only staging, BIOT achieves the best result with a macro~F1 of 0.7237, followed by LaBraM (0.6835) and SleepFM from scratch (0.6582). Across the five models capable of ECG-only staging, switching from EEG to ECG costs between 0.2798 (MOMENT) and 0.4151 (BIOT) macro~F1, averaging 0.3531, while cutting the raw channel data rate to a third. Adding ECG to EEG provides no gain for most models. These results show that EEG carries most of the sleep-staging signal, quantify the consistent accuracy cost of the wearable-compatible alternative, and demonstrate that sleep-relevant pretraining transfers well to MESA. Because sleep staging is itself an intermediate step toward downstream applications such as dementia and other disease risk prediction, understanding the accuracy-versus-deployability trade-off is a prerequisite for building sleep-monitoring pipelines that can run outside the clinic.
\end{abstract}

\section{Introduction}

Sleep is a fundamental physiological process that supports memory consolidation, metabolic regulation, cardiovascular health, and immune function~\cite{irfan2025novel,irfan2025multidomain,irfan2025smart}. Insufficient or disrupted sleep is associated with cardiovascular disease, cognitive decline, and increased mortality risk, making accurate, scalable sleep assessment a public health concern rather than only a clinical one~\cite{thapa2026sleepfm}. Polysomnography (PSG) is the clinical gold standard for sleep assessment~\cite{thapa2026sleepfm}. A PSG recording captures EEG, electrooculography (EOG), electromyography (EMG), ECG, and respiratory signals in parallel overnight, and a trained technician scores each 30-second epoch into one of five stages: Wake; three progressively deeper stages of non-REM sleep (N1, N2, N3); or rapid eye movement (REM) sleep, following the American Academy of Sleep Medicine (AASM) criteria. This process is accurate but has real drawbacks: PSG requires a clinic visit, multiple wired sensors attached to the body, and several hours of expert manual scoring per recording, which limits how often and how widely it can be used. Manual scoring is also not perfectly consistent between experts, since inter-rater agreement on staging typically falls short of full agreement even among trained scorers~\cite{lee2022interrater}.

Automating sleep staging with machine learning addresses the scoring bottleneck directly: a trained model can score a full night in seconds rather than hours, at consistent, reproducible criteria. This is a precondition for using sleep staging at a scale that PSG alone cannot support, including continuous, long-term, at-home monitoring, and for downstream applications that depend on it. Sleep-derived signals have been used to predict conditions well beyond sleep disorders themselves, including dementia and other disease risks~\cite{thapa2026sleepfm}, which motivates broader work on dementia-relevant biomarkers from sleep. Automated, minimal-signal staging that can run outside the clinic is a prerequisite for using sleep as a biomarker source at scale, not merely a technical convenience.

\begin{figure}[t] 
\centering 
\includegraphics[width=\linewidth]{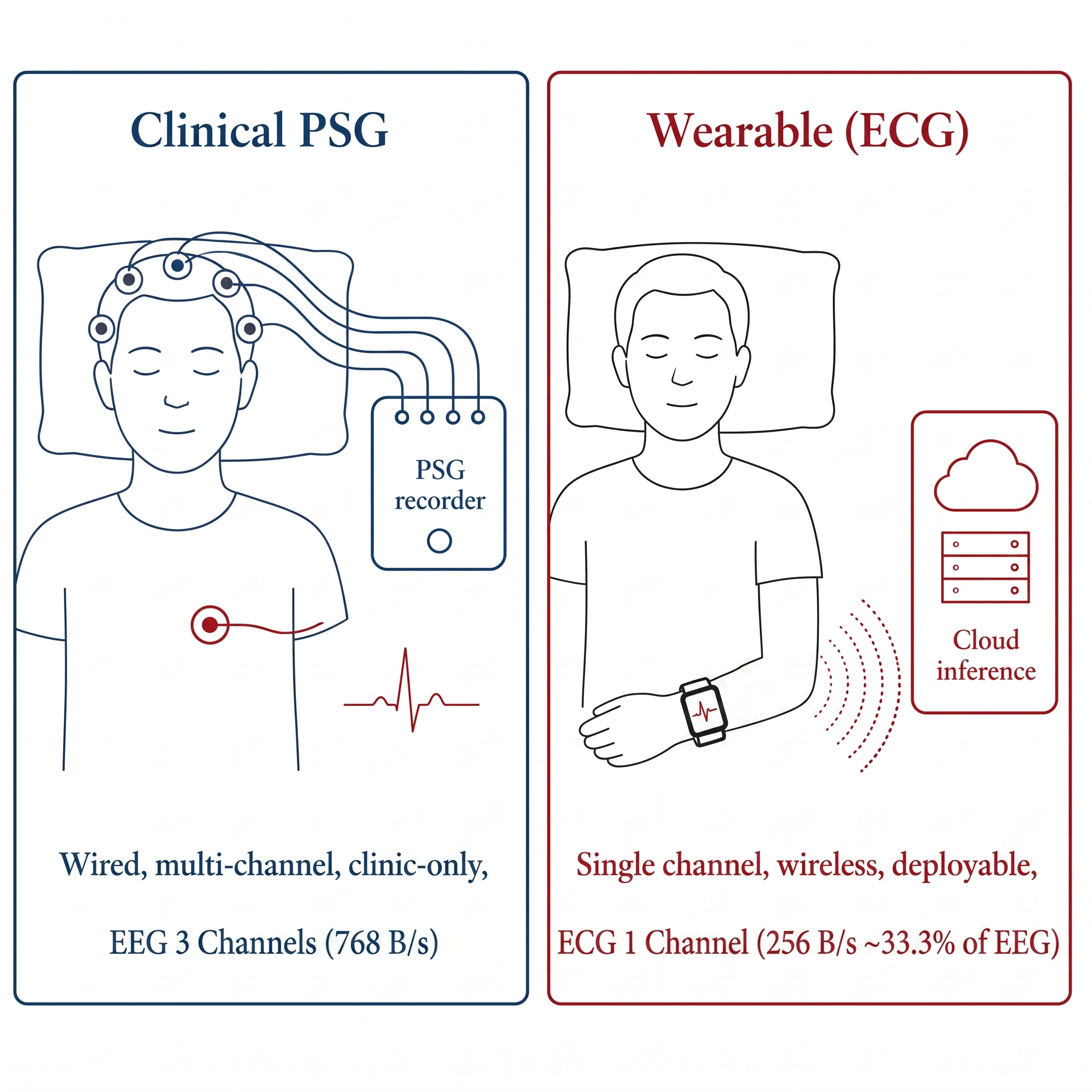} 
\caption{Clinical PSG montage compared to a single wearable device.} \label{fig:motivation} 
\end{figure}

A practical question follows directly from PSG's drawbacks: how much does each individual signal in the full montage actually contribute to staging accuracy? ECG and other cardiac or motion signals are already captured by consumer wearables, whereas EEG requires scalp electrodes. If ECG alone could match EEG-based performance, it would open the door to sleep staging outside the clinic (Figure~\ref{fig:motivation}).

Wearable devices make continuous, long-term sleep sensing feasible outside the laboratory~\cite{mogavero2025wearable}. But deployment constraints are often underemphasized in model design. EEG requires scalp electrodes and is largely confined to clinical or laboratory settings, whereas ECG is already captured by consumer wearables such as chest straps, wristbands, and rings. If sleep staging models trained on clinical PSG can be reused with only ECG, or with EEG downgraded to a minimal 3-channel montage, that changes what is deployable outside the clinic. This motivates a specific question: not just whether EEG beats ECG, but by how much. What is the accuracy cost of choosing the wearable-compatible signal (ECG) over the clinical one (EEG), and does that cost hold consistently across different model architectures and pretraining regimes? Related work on minimal-sensing approaches has shown that a single physiological channel can support meaningful classification when paired with appropriate feature design, motivating the same question for sleep staging with modern foundation models~\cite{irfan2025minimal}.

Foundation models for biosignals have recently shown strong results on sleep staging~\cite{thapa2026sleepfm,yang2023biot,carter2025wav2sleep,shuai2026osf}. Most papers report full-modality performance without systematically comparing which single signal a model actually needs. Lee et al.~\cite{lee2025large} find that large brainwave foundation models achieve only marginal improvements over traditional architectures. We ask a related question specific to sleep: across models with very different architectures and pretraining, how much macro~F1 is lost by staging from ECG alone instead of EEG alone, and does adding ECG to EEG ever help enough to justify the extra channel?

We reproduce SleepFM~\cite{thapa2026sleepfm} on the MESA Sleep Study~\cite{chen2015mesa}, train an encoder from scratch on MESA to remove encoder-level leakage from the published checkpoint, and compare against BIOT~\cite{yang2023biot} and LaBraM~\cite{jiang2024labram} (pretrained foundation models), MOMENT~\cite{goswami2024moment} (general time-series foundation model), a from-scratch ViT-B based on SensorLM~\cite{zhang2025sensorlm} (no pretraining), and YASA~\cite{vallat2021yasa} (classical baseline). We restrict the comparison to the three signal conditions available across all six models: EEG, ECG, and EEG+ECG.

The remainder of this paper is organised as follows. Section~\ref{sec:relatedwork} reviews related work on sleep-staging foundation models and wearable/edge sensing. Section~\ref{sec:methods} describes the MESA dataset, our preprocessing pipeline, the signal conditions and their raw data rates, and the six models we compare. Section~\ref{sec:results} reports macro~F1 across EEG, ECG, and EEG+ECG for all six models. Section~\ref{sec:discussion} discusses what these results imply for wearable and edge-cloud sleep staging, and Section~\ref{sec:limitations} states the limitations of the study, and Section VII provides the conclusion of the study. 

\FloatBarrier

\section{Related Work}\label{sec:relatedwork}

Foundation models for physiological time series generally follow one of three strategies: domain-general pretraining on time series drawn from many application domains, domain-specific pretraining on large-scale clinical EEG corpora, and no pretraining at all, with the model trained directly on the target dataset. Our comparison includes a representative of each strategy, separating two effects that single-model papers usually confound: how much a foundation model's pretraining domain determines its downstream performance, and how much is instead attributable to its raw architectural capacity.

BIOT~\cite{yang2023biot} and LaBraM~\cite{jiang2024labram} both represent the domain-specific EEG-pretrained strategy, though neither was pretrained on ECG. BIOT's EEG-SHHS+PREST checkpoint, which we fine-tune here, was pretrained on the C3/A2 and C4/A1 EEG channels from the Sleep Heart Health Study (SHHS) together with EEG data from PREST, using Short-Time Fourier Transform (STFT) tokenisation to normalise across the variable channel counts and recording hardware found across different EEG datasets. LaBraM takes a related but narrower approach, pretraining specifically on EEG using a vector-quantised tokeniser. Neither BIOT's SHHS+PREST checkpoint nor LaBraM was exposed to ECG during pretraining, so both models must learn to use the ECG channel entirely at fine-tuning time when we apply them to EEG+ECG input in our comparison.

MOMENT~\cite{goswami2024moment} represents the domain-general strategy: it is pretrained on the Time-series Pile, a large-scale public corpus of time series spanning many application domains (e.g.\ energy, healthcare, transportation) with no biosignal-specific supervision, and we use it with a frozen encoder, so all task adaptation happens in a linear head fit on top of general-purpose embeddings. SensorLM's architecture~\cite{zhang2025sensorlm} and SleepFM~\cite{thapa2026sleepfm}, as we use them, sit at the no-pretraining end of this spectrum. SensorLM's original released weights are not available to us, so we reimplement its ViT-B sensor encoder and train it directly on MESA. SleepFM's published encoder is excluded from our comparison because MESA is part of its own pretraining data, so we train a separate encoder from scratch on the same 270-subject training split used for every other model. YASA~\cite{vallat2021yasa} sits outside this axis entirely: it is a gradient-boosted-tree pipeline over handcrafted PSG features, applied zero-shot, and it establishes a reference point for performance without any learned representation.

Other recent work on sleep-staging foundation models, including wav2sleep~\cite{carter2025wav2sleep} and OSF~\cite{shuai2026osf}, reports strong full-modality results but, to our knowledge, does not provide a systematic single-signal ablation comparing EEG-only versus ECG-only staging across multiple independently pretrained architectures. Lee et al.~\cite{lee2025large} evaluate large brainwave foundation models against smaller task-specific architectures and find only marginal gains from scale, a finding our EEG-only results are broadly consistent with: SensorLM, trained from scratch with no pretraining, comes within 0.0571 macro~F1 of LaBraM, a pretrained EEG foundation model.

The broader motivation for this comparison sits in the growing literature on wearable and edge-deployed physiological sensing. Mogavero et al.~\cite{mogavero2025wearable} trace the shift in wearable sleep monitoring from movement-based actigraphy toward multi-sensor, AI-driven systems, and note the growing role of edge computing, specifically in automated sleep staging. Separately, minimal-sensing work such as Irfan et al.~\cite{irfan2025minimal} shows that a single physiological channel, paired with appropriate feature design, can support a meaningful classification task under the battery and data-transmission constraints of a wearable device, though that result is for emotion classification from ECG rather than sleep staging. Our work asks whether an analogous single-channel reduction holds specifically for sleep staging and whether it holds consistently across several independently pretrained model architectures rather than just one.

Prior work specifically evaluated on MESA spans a wide range of signal counts without holding the architecture constant across that range. At the high end, Fox et al.~\cite{fox2024pftsleep} train PFTSleep, a foundational transformer over 7 PSG channels (brain, movement, cardiac, oxygen, and respiratory signals), on SHHS and use MESA purely as an external test set, reporting 0.5697 macro~F1. Zan and Yildiz~\cite{zan2023fullsleepnet} instead use a single EEG channel on both SHHS and MESA, reporting 0.761 macro~F1 and 0.829 accuracy on MESA with a multi-task architecture combining a convolutional neural network (CNN), a recurrent neural network (RNN), and attention that jointly scores arousals and sleep stages. At the ECG end, Sridhar et al.~\cite{sridhar2020heartrate} train on SHHS and MESA using only the instantaneous heart rate extracted from ECG, reporting 0.77 accuracy on a 4-class staging task, and Attia et al.~\cite{attia2025sleepppgnet2} report a similar single-signal approach using photoplethysmography (PPG) instead, evaluated on MESA among other cohorts. These results show that single-channel and full-montage approaches have been tried separately on MESA, each with its own architecture, but not systematically within a single architecture family, holding the model constant while varying only the channel count. That is the specific comparison our EEG/ECG/EEG+ECG evaluation is designed to make, across six architecturally distinct models rather than one.

\FloatBarrier

\section{Methods}\label{sec:methods}

\begin{figure*}[t]
\centering
\includegraphics[width=0.92\textwidth]{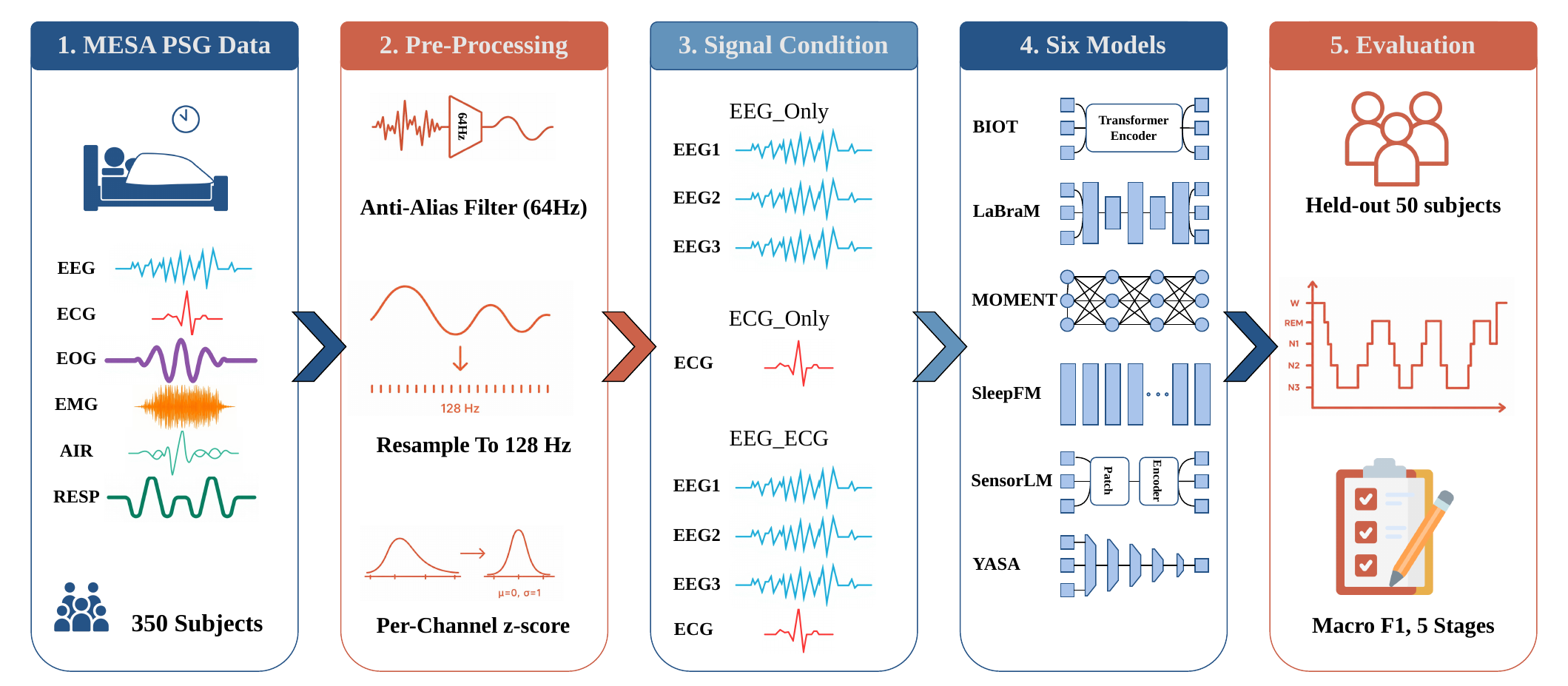}
\caption{Overview of the evaluation pipeline.}
\label{fig:pipeline}
\end{figure*}

Figure~\ref{fig:pipeline} outlines the full evaluation pipeline described in this section. This study does not modify the architecture of any of the evaluated models. BIOT, LaBraM, MOMENT, and SleepFM are used with their original, published architectures, either fine-tuned or with a frozen encoder, as described for each model below. SensorLM is the one exception in implementation, not architecture: because no released weights exist, we reimplement its published sensor-encoder architecture in PyTorch and train it from scratch, but the architecture itself follows the original specification. The three signal conditions we evaluate, EEG\_ONLY, ECG\_ONLY, and EEG+ECG, constitute an input ablation: the same fixed architecture and training procedure per model is evaluated under a controlled change to only the input channels, which isolates the contribution of each signal from any change in model capacity or training regime.

\subsection{Dataset}\label{sec:dataset}

We use the MESA Sleep Study~\cite{chen2015mesa}, a publicly available PSG dataset from the National Sleep Research Resource (NSRR). The full MESA Sleep cohort includes PSG recordings from over 2,000 participants; we use a fixed subsample of 350 subjects due to memory problem w had initially, consistent with a pilot-scale evaluation across six independently trained or fine-tuned models. We discuss this choice further in Section~\ref{sec:limitations}. Labels follow the AASM 5-class scheme (Wake, N1, N2, N3, REM), one label per 30-second epoch; the raw annotation files store stages as variable-length events, which we expand into per-epoch rows to match this scheme.

We define two signal groups used across all models: \textbf{EEG\_ONLY} (EEG1+EEG2+EEG3, MESA's raw channel labels for its three scalp EEG derivations, mapped to standard 10-20 electrode positions in Section~\ref{sec:baselines}) and \textbf{EKG} (one ECG channel), combined as \textbf{EEG+ECG}. MESA includes additional PSG channels (EOG, respiratory, EMG), which we do not use here since not every model in our comparison has a corresponding checkpoint or configuration for those signals. Raw MESA European Data Format (EDF) signals are recorded with 16-bit signed Analogue-to-Digital Converter (ADC) resolution. All channels are converted to Hierarchical Data Format version 5 (HDF5) with \texttt{mne.io.read\_raw\_edf} (the EDF-loading function from MNE-Python, a standard open-source library for biosignal data), a 4th-order Butterworth low-pass anti-aliasing filter (cutoff at the Nyquist frequency of the target rate, 64\,Hz, applied with \texttt{filtfilt}, SciPy's zero-phase forward-backwards filtering function), linear-interpolation resampling to 128\,Hz, and per-channel z-score normalisation. Subjects are split with a fixed random seed (42) into 270 training, 30 validation, and 50 held-out test subjects. %This partition is generated using a 10-fold splitting scheme, but we use only the first fold (fold\_0) throughout this study; we do not train or evaluate on the remaining nine folds. This same fold\_0 split is used for all models we fine-tune, and it is not stratified by demographics or disease severity.

\subsection{Signal Configurations and Data Volume}

We focus on three signal conditions available across all six models: EEG\_ONLY (EEG1+EEG2+EEG3, clinic-grade, scalp electrodes), EKG (one ECG channel, compatible with consumer wearables), and their combination. Because every channel is resampled to a common 128\,Hz rate and the original MESA recordings use 16-bit ADC resolution, the raw data rate of a channel set is 2 bytes $\times$ 128\,Hz $\times$ number of channels. Table~\ref{tab:datavol} reports this rate for each condition, relative to EEG\_ONLY as the clinical baseline. We report only this raw-channel data rate; we do not have measured power consumption or wireless bandwidth figures for any device, so we make no claims about battery life or transmission bandwidth.

\begin{table}[t]
\caption{Raw data rate by signal condition, at the 128\,Hz resample rate and 16-bit ADC resolution described in Section~\ref{sec:dataset}. \% is relative to EEG\_ONLY.}
\label{tab:datavol}
\centering
\normalsize
\renewcommand{\arraystretch}{1.25}
\setlength{\tabcolsep}{3pt}
\begin{tabular*}{\linewidth}{@{\extracolsep{\fill}}lccc}
\toprule
Configuration & Channels & Rate (B/s) & \% of EEG\_ONLY \\
\midrule
EEG+ECG & 4 & 1024 & 133.3\% \\
EEG\_ONLY & 3 & 768 & 100.0\% \\
ECG\_ONLY & 1 & 256 & 33.3\% \\
\bottomrule
\end{tabular*}
\end{table}

\subsection{SleepFM}

 SleepFM~\cite{thapa2026sleepfm} uses a SetTransformer encoder (embed\_dim~128, 8~heads, 6~layers) processing 5-second windows at 128\,Hz and a 2-layer bidirectional long short-term memory (LSTM) staging head (dropout~0.3, 5~classes). It was pretrained on over 585,000 hours of PSG from approximately 65,000 participants across Stanford Sleep Clinic, BioSerenity, MESA, and MrOS. %Because MESA was included in the pretraining data, the published checkpoint exhibits encoder-level leakage on our evaluation set.
 We train an EEG\_ONLY encoder from scratch on 270 MESA subjects using leave-one-out contrastive pretraining, for the EEG/ECG comparison. Contrastive pretraining uses stochastic gradient descent (SGD, momentum~0.9) with a step learning-rate (lr) schedule (initial lr~0.001, step every 2 epochs, $\gamma$~0.1) and a batch size of 128. Fine-tuning for staging uses Adam (lr~0.0001), batch size~16 with 16-step gradient accumulation (effective batch~256), up to 500 epochs with early-stopping patience~50, and an LSTM-based staging head (embed\_dim~128, 4~heads, 2~layers, 5~classes, dropout~0.3; 1.83M trainable parameters).

\subsection{Baseline Models}\label{sec:baselines}

\textbf{BIOT}~\cite{yang2023biot} is a biosignal foundation model. The EEG-SHHS+PREST checkpoint we use was pretrained on two EEG channels (C3/A2, C4/A1) from SHHS and EEG data from PREST, using STFT tokenisation and a linear-attention transformer; this checkpoint was not exposed to ECG during pretraining. We fine-tune end-to-end using this EEG-SHHS+PREST-18-channels checkpoint with AdamW (lr~0.0001), batch size~16, and early-stopping patience~20. Input is resampled from 128\,Hz to 200\,Hz (6000 samples per 30-second epoch). The classification head uses embed\_size~256, 8~heads, depth~4, with STFT parameters n\_fft~200 and hop\_length~100. We use a cross-entropy loss with inverse-frequency class weights.

\textbf{MOMENT}~\cite{goswami2024moment} is a general time-series foundation model (MOMENT-1-large checkpoint, 341M parameters) pretrained via masked reconstruction on the Time-series Pile. We freeze both the encoder and the embedder and train only a linear classification head (\texttt{Linear(n\_channels$\times$1024, 5)}, concatenation reduction) with AdamW (lr~0.0001), batch size~8, and early-stopping patience~20.

\textbf{LaBraM}~\cite{jiang2024labram} is an EEG foundation model pretrained on large-scale clinical EEG via vector-quantised neural spectrum prediction. We map MESA channels to 10-20 electrode names verified against the MESA PSG manual (EEG1$\to$Fz, EEG2$\to$Oz, EEG3$\to$C4, EKG$\to$T9) and fine-tune end-to-end from the \texttt{labram-base} checkpoint with AdamW (lr~0.0001), batch size~16, and early-stopping patience~20. Each 30-second epoch is split into two 15-second chunks of 15 one-second patches each, with per-chunk embeddings averaged; signal amplitude is divided by 100 following the LaBraM convention. We use a cross-entropy loss with class weights.

\textbf{SensorLM}~\cite{zhang2025sensorlm} is a wearable sensor-language model with no released weights. We reimplement its sensor encoder architecture (ViT-B: depth~12, embed\_dim~768, mlp\_dim~3072, heads~12, multihead attention pooling (MAP), patch size 0.5\,s $\times$ 1~channel, 60 time patches per channel, factored positional embedding; 92.2M parameters) in PyTorch and train from scratch on MESA with AdamW (lr~0.0001, weight\_decay~0.0001, gradient clipping~1.0), batch size~64, and a cross-entropy loss with inverse-frequency class weights. This is not a pretrained model.

\textbf{YASA}~\cite{vallat2021yasa} is a classical sleep staging tool based on handcrafted PSG features and gradient boosting. Its feature set is built from EEG-specific spectral characteristics, so it cannot process ECG alone without redesigning that feature set, which would count as a model modification and fall outside the scope of this comparison. Applied zero-shot.

\subsection{Evaluation}

All models except the YASA are evaluated on the same fixed held-out split: 50 test subjects (49 with labels, approximately 60,550 epochs). We report macro-averaged F1 across all 5 classes (the unweighted mean of the per-class F1 scores for Wake, N1, N2, N3, and REM), alongside overall accuracy (Acc), the fraction of epochs correctly classified across all stages. Macro-averaging weights each stage equally regardless of how often it occurs in the test set, which matters here because Wake and N2 together account for most epochs while N1 is comparatively rare; a metric weighted by class frequency would understate exactly the N1 degradation we report in Sections~4 and~5. We evaluate each model once on the fixed held-out split rather than across multiple folds or random seeds, so the F1 values we report do not include confidence intervals; we return to this in Section~\ref{sec:limitations}.

\FloatBarrier

\section{Results}\label{sec:results}

\subsection{Multi-Model Comparison}

\begin{figure}[t]
\centering
\includegraphics[width=\linewidth]{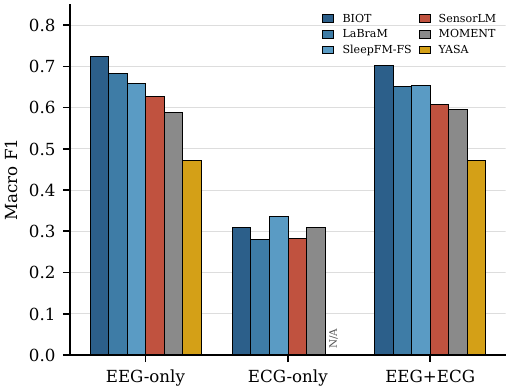}
\caption{Macro F1 by model across the three signal conditions. YASA is omitted from ECG-only, which it cannot run.}\label{fig:bars}
\end{figure}

Tables~\ref{tab:eeg}--\ref{tab:eegecg} compare all models on the three signal conditions using the same held-out test split. Figure~\ref{fig:bars} summarises macro~F1 across all three conditions for all six models.

\paragraph{EEG-only} BIOT leads at 0.7237 macro~F1 and produces the best per-class F1 for Wake, N1, N3, and REM, though not N2, where SleepFM from-scratch is best (0.7653) despite ranking third overall. No single model dominates every stage, even though one model dominates the macro average. LaBraM (0.6835) benefits from EEG-specific pretraining. SleepFM from-scratch (0.6582) and the SensorLM architecture trained from scratch (0.6264) are close, suggesting the ViT-B architecture is capable even without sleep-specific pretraining, though the 0.0571 gap to LaBraM still favours pretraining. MOMENT (0.5894, frozen) is the weakest trained model, consistent with adapting only a linear head on general-purpose embeddings rather than fine-tuning end-to-end. YASA (0.4720, zero-shot) underperforms all trained models, particularly on N3, where it is the only model below 0.6 (Table~\ref{tab:eeg}).

\paragraph{ECG-only} All models struggle, and the ranking changes substantially from EEG-only. SleepFM from-scratch leads at 0.3353, followed by MOMENT (0.3096), BIOT (0.3086), SensorLM (0.2821), and LaBraM (0.2803). BIOT and LaBraM, the two strongest EEG-only models, fall to third and last on ECG-only, while MOMENT rises from fifth to second. MOMENT also produces the best N1 (0.1850) and N2 (0.4467) F1 of any model on ECG-only, despite being the weakest trained model on EEG-only; its general-purpose, frozen embeddings appear to transfer to ECG's simpler waveform structure at least as well as embeddings from models pretrained specifically on EEG or PSG. N1~F1 is 0.0000 for SleepFM-FS and near-zero for every other model. N3~F1 stays below 0.22 across the board. (YASA is excluded from this comparison entirely, since it requires an EEG channel by design, as noted in Section~\ref{sec:baselines}.).

\begin{figure*}[t]
\centering
\includegraphics[width=0.85\textwidth]{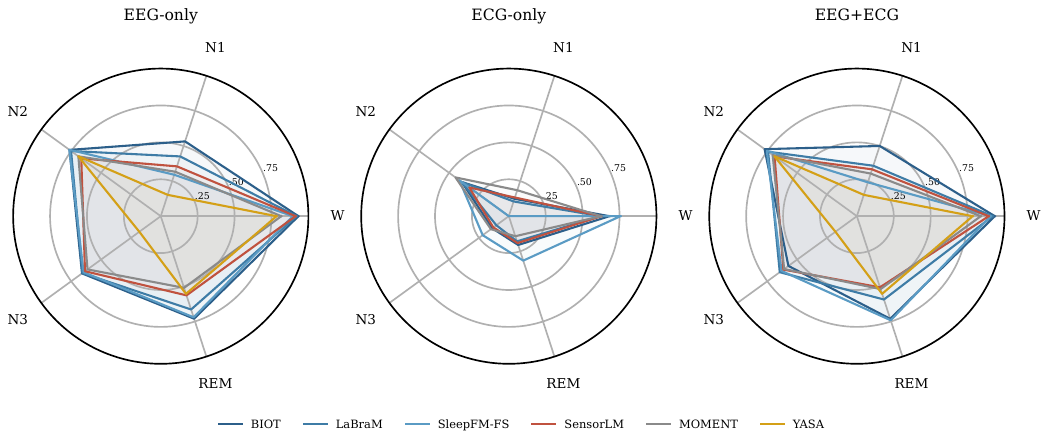}
\caption{Per-stage F1 by model, one radar per signal condition. Spokes are Wake/N1/N2/N3/REM F1; each line is one model. YASA cannot run ECG-only and is omitted from the centre panel.}
\label{fig:radar}
\end{figure*}

\paragraph{EEG+ECG} Adding ECG hurts most models. BIOT drops from 0.7237 to 0.7023, LaBraM from 0.6835 to 0.6524, and SleepFM from scratch from 0.6582 to 0.6529. MOMENT is the exception: F1 rises from 0.5894 to 0.5953. For BIOT, this net loss is driven almost entirely by N3, which drops from 0.6594 to 0.5733, a decrease of 0.0861, while Wake, N2, and REM each change by less than 0.01 in either direction; adding ECG does not degrade BIOT uniformly across stages, it concentrates the damage in N3. YASA results are identical to EEG-only since it ignores the ECG channel. Figure~\ref{fig:radar} summarises the per-stage F1 breakdown discussed above for all six models across the three signal conditions.

\begin{table}[t]
\caption{EEG-only staging, all models, held-out 50 test. Bold marks the best score in each column.}\label{tab:eeg}
\centering
\footnotesize
\renewcommand{\arraystretch}{1.3}
\setlength{\tabcolsep}{1.5pt}
\begin{tabular*}{\linewidth}{@{\extracolsep{\fill}}lccccccc}
\toprule
Model & F1 & Acc & W & N1 & N2 & N3 & REM \\
\midrule
BIOT & \textbf{0.7237} & \textbf{0.8007} & \textbf{0.9325} & \textbf{0.5327} & 0.7639 & \textbf{0.6594} & \textbf{0.7301} \\
LaBraM & 0.6835 & 0.7737 & 0.9203 & 0.4257 & 0.7511 & 0.6557 & 0.6647 \\
SleepFM-FS & 0.6582 & 0.7757 & 0.8941 & 0.2952 & \textbf{0.7653} & 0.6464 & 0.7201 \\
SensorLM & 0.6264 & 0.7235 & 0.9132 & 0.3549 & 0.6657 & 0.6344 & 0.5639 \\
MOMENT & 0.5894 & 0.6745 & 0.8163 & 0.3166 & 0.6900 & 0.6134 & 0.5105 \\
YASA & 0.4720 & 0.6687 & 0.7843 & 0.1519 & 0.6913 & 0.1795 & 0.5531 \\
\bottomrule
\end{tabular*}
\end{table}

\begin{table}[t]
\caption{ECG-only staging, all models. Bold marks the best score in each column. YASA = N/A.}\label{tab:ecg}
\centering
\footnotesize
\renewcommand{\arraystretch}{1.3}
\setlength{\tabcolsep}{1.5pt}
\begin{tabular*}{\linewidth}{@{\extracolsep{\fill}}lccccccc}
\toprule
Model & F1 & Acc & W & N1 & N2 & N3 & REM \\
\midrule
SleepFM-FS & \textbf{0.3353} & \textbf{0.5268} & \textbf{0.7566} & 0.0000 & 0.4148 & \textbf{0.2188} & \textbf{0.3165} \\
MOMENT & 0.3096 & 0.4191 & 0.6239 & \textbf{0.1850} & \textbf{0.4467} & 0.1498 & 0.1427 \\
BIOT & 0.3086 & 0.4302 & 0.6815 & 0.1239 & 0.3933 & 0.1391 & 0.2051 \\
SensorLM & 0.2821 & 0.3695 & 0.6257 & 0.1330 & 0.3307 & 0.1296 & 0.1914 \\
LaBraM & 0.2803 & 0.4092 & 0.6211 & 0.1049 & 0.3920 & 0.1045 & 0.1792 \\
YASA & \multicolumn{7}{c}{N/A (requires EEG)} \\
\bottomrule
\end{tabular*}
\end{table}

\begin{table}[t]
\caption{EEG+ECG staging, all models. Bold marks the best score in each column. YASA uses EEG1 only.}\label{tab:eegecg}
\centering
\footnotesize
\renewcommand{\arraystretch}{1.3}
\setlength{\tabcolsep}{1.5pt}
\begin{tabular*}{\linewidth}{@{\extracolsep{\fill}}lccccccc}
\toprule
Model & F1 & Acc & W & N1 & N2 & N3 & REM \\
\midrule
BIOT & \textbf{0.7023} & \textbf{0.7975} & \textbf{0.9345} & \textbf{0.5001} & \textbf{0.7724} & 0.5733 & 0.7311 \\
SleepFM-FS & 0.6529 & 0.7765 & 0.9069 & 0.2411 & 0.7542 & 0.6309 & \textbf{0.7415} \\
LaBraM & 0.6524 & 0.7566 & 0.9203 & 0.3594 & 0.7431 & \textbf{0.6457} & 0.5935 \\
SensorLM & 0.6077 & 0.7062 & 0.8935 & 0.3337 & 0.6891 & 0.6155 & 0.5068 \\
MOMENT & 0.5953 & 0.6918 & 0.8395 & 0.3035 & 0.7053 & 0.6089 & 0.5194 \\
YASA & 0.4720 & 0.6687 & 0.7843 & 0.1519 & 0.6913 & 0.1795 & 0.5531 \\
\bottomrule
\end{tabular*}
\end{table}

\subsection{Per-Subject Analysis}
\label{sec:persubject}

The comparisons above report a single aggregate macro~F1 per model per condition across the full held-out test set. This masks two questions worth answering directly: does performance vary meaningfully from one subject to the next, and are the differences between models real or within the range of subject-to-subject noise? We answer both using per-subject predictions for BIOT, SleepFM from-scratch, LaBraM, SensorLM, MOMENT, and YASA (EEG-only and EEG+ECG only; YASA cannot run ECG-only).

\begin{figure*}[t]
\centering
\includegraphics[width=0.85\textwidth]{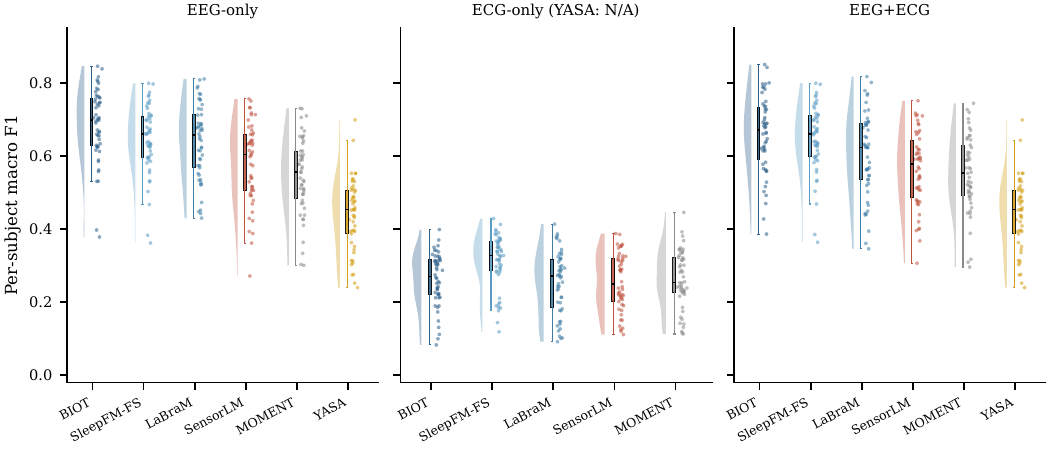}
\caption{Per-subject macro F1 distribution for BIOT, SleepFM from-scratch, LaBraM, SensorLM, MOMENT, and YASA, one panel per signal condition. YASA cannot run ECG-only and is omitted from that panel. Points are individual subjects.}
\label{fig:violin}
\end{figure*}

Figure~\ref{fig:violin} shows the full per-subject distribution behind each comparison. On EEG-only, BIOT wins on 39 of 50 subjects, LaBraM on 7, YASA and SensorLM on 2 each, and SleepFM from-scratch and MOMENT on none; nine of the ten pairwise comparisons among BIOT, SleepFM from-scratch, LaBraM, SensorLM, and MOMENT are statistically significant (Wilcoxon signed-rank, $p<0.0001$ in every case except SensorLM vs.\ MOMENT at $p=0.0004$), the one exception being SleepFM from-scratch vs.\ LaBraM ($p=0.856$, not significant). On EEG+ECG, BIOT wins 25 of 50, SleepFM from-scratch 12, LaBraM 8, YASA 3, and SensorLM and MOMENT 1 each; eight of ten pairwise comparisons are significant, the exceptions being BIOT vs.\ SleepFM from-scratch ($p=0.138$) and SensorLM vs.\ MOMENT ($p=0.291$). YASA is shown for visual comparison in both panels but was not included in this pairwise testing.

On ECG-only, where YASA cannot run, the picture changes once SleepFM from-scratch is included. It wins 31 of 50 subjects outright, more than the other four models combined (BIOT 4, LaBraM 3, SensorLM 3, MOMENT 9), and is significantly better than every one of them ($p<0.0001$ in all four comparisons). Among BIOT, LaBraM, SensorLM, and MOMENT, however, all six pairwise comparisons are non-significant ($p$ ranging from 0.270 to 0.977). So the noise-floor pattern still holds for four of the five models capable of ECG-only staging, but SleepFM from-scratch is a genuine exception: its higher aggregate ECG-only macro~F1 (0.3353, the best of any model) reflects a real, subject-level advantage rather than sitting within the same noise band as the rest. Across the two conditions in which it can run, YASA has the lowest median per-subject F1 of all six models (0.4527, identical in both conditions since it ignores ECG), consistent with its role as a classical, zero-shot baseline rather than a learned representation.

\section{Discussion}\label{sec:discussion}

This study evaluates six sleep-staging models under a controlled signal reduction, from a full EEG montage down to a single ECG channel, to quantify what a wearable-only deployment would actually cost in staging accuracy. The comparison spans models built on different pretraining strategies, EEG-pretrained, general time-series, and from-scratch, plus one classical non-learned baseline, all evaluated on the same held-out MESA test split under three signal conditions: EEG, ECG, and EEG+ECG. The main finding is that EEG-only staging is close to or better than EEG+ECG for nearly every model, that ECG alone costs a substantial and fairly consistent amount of accuracy relative to EEG alone, and that this pattern holds regardless of how a given model was pretrained.

\paragraph{Strengths of this comparison.} Three aspects of the design support these conclusions. First, no model's architecture is modified from its original, published form, so differences between models reflect the models themselves rather than any intervention on our part. Second, every model is evaluated on the same three signal conditions and the same fixed held-out test split, so the EEG-to-ECG comparison is a controlled input ablation rather than a comparison across differently constructed experiments. Third, the accuracy cost of switching from EEG to ECG replicates across five architecturally distinct pretraining strategies rather than resting on a single model, which is why we treat it as a property of the ECG signal rather than of any one architecture. The per-subject analysis in Section~\ref{sec:persubject} supports this directly: across BIOT, LaBraM, MOMENT, and SensorLM, every one of the six pairwise comparisons between models is statistically non-significant on ECG-only ($p$ ranging from 0.270 to 0.977), while the same pairs are overwhelmingly significant on EEG-only and EEG+ECG. Model identity stops predicting performance specifically when the EEG is removed, providing stronger evidence than aggregate consistency alone. SleepFM from-scratch is the one exception to this specific pattern, performing significantly better than all four other models on ECG-only; we discussed this in Section~\ref{sec:persubject}.

\paragraph{EEG is the dominant signal.} For four of six models, EEG-only staging outperforms EEG+ECG; YASA ties exactly since it ignores the ECG channel, and MOMENT is the one model where EEG+ECG wins outright. BIOT and LaBraM, whose pretraining checkpoints were never exposed to ECG, both drop when ECG is added: BIOT by 0.0214 macro~F1 and LaBraM by 0.0311. SleepFM from-scratch and SensorLM, which have no pretraining at all, drop by 0.0053 and 0.0187, respectively. MOMENT is the only exception, gaining 0.0059 macro~F1 with ECG added. Unlike BIOT and LaBraM, MOMENT's Time-Series Pile pretraining corpus does include ECG recordings from healthcare-domain datasets, though not paired with EEG in a sleep-staging context; this may explain why MOMENT can make some use of the added channel, where BIOT and LaBraM, encountering ECG for the first time at fine-tuning, cannot. Because MOMENT's encoder is frozen and only the linear head is trained, this gain is still small enough that it may equally be a product of the extra input dimensionality reaching a slightly better linear fit, rather than genuinely complementary information extracted from ECG. Across the four models that do drop, regardless of prior ECG exposure, ECG mostly does not add discriminative information beyond what EEG already captures.

\paragraph{ECG alone is not enough.} N1~F1 is effectively zero for every model on ECG-only input. N3~F1 stays below 0.22. These stages require fine-grained frequency analysis of brain activity. Cardiac signals, averaged over coarse time windows, reflect the autonomic state and cannot resolve the subtle EEG signatures that define N1 and N3.

\begin{figure*}[t]
\centering
\begin{subfigure}[b]{0.48\textwidth}
\centering
\includegraphics[width=\linewidth]{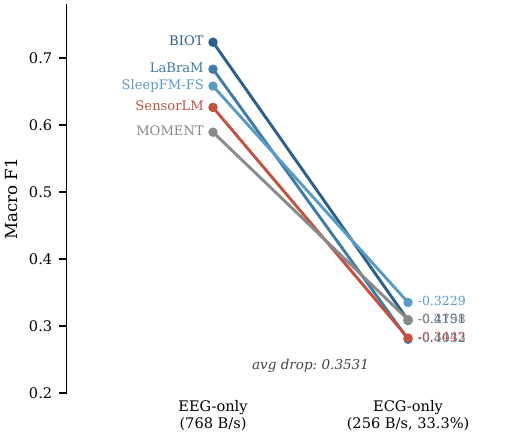}
\caption{}
\label{fig:slope}
\end{subfigure}
\hfill
\begin{subfigure}[b]{0.48\textwidth}
\centering
\includegraphics[width=\linewidth]{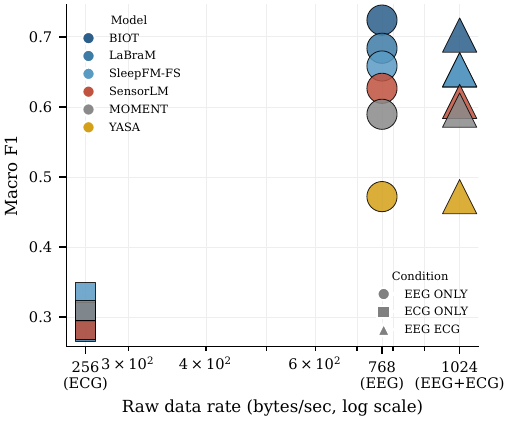}
\caption{}
\label{fig:scatter}
\end{subfigure}
\caption{The deployment cost of choosing ECG over EEG. (a) EEG-only vs.\ ECG-only macro F1 for the five models capable of ECG-only staging: direction is consistent, magnitude varies. (b) Macro F1 versus raw channel data rate (log scale), all models, all three signal conditions; marker shape denotes condition, marker size scales with channel count.}
\label{fig:deployment}
\end{figure*}

\paragraph{Implications for longitudinal home monitoring.} The motivating use case for this comparison is repeated, long-term sleep monitoring outside the clinic, which is exactly the setting where the EEG requirement is hardest to satisfy: multi-night scalp EEG is not something most people will tolerate at home over months. Our results do not show that ECG is an adequate substitute in absolute terms. The weakest ECG-only results, LaBraM at 0.2803 and SensorLM at 0.2821 macro~F1, are not close to clinical-grade staging, and even the strongest, SleepFM from-scratch at 0.3353, trails EEG-only by 0.3229. What the results do show is that this gap is consistent enough across five architecturally different models, spanning EEG-pretrained, general time-series, and from-scratch strategies, to treat it as a property of the ECG signal itself rather than of any one model's training. That distinction matters for how future work is directed: it argues for research into what ECG-derived features or auxiliary signals could close the gap, rather than for testing more foundation model architectures on the same single-channel ECG input.

\paragraph{Sleep-relevant pretraining helps.} The gap between BIOT (0.7237, EEG-pretrained) and SensorLM (0.6264, from-scratch) on EEG-only is 0.0973 macro~F1. MOMENT (0.5894, general time-series, frozen, 341M parameters) is weaker than SensorLM (92.2M parameters) despite having roughly 3.7$\times$ more parameters. Domain-specific pretraining matters more than model size. This aligns with Lee et al.~\cite{lee2025large}, who show that large brainwave foundation models provide only marginal gains over smaller task-specific architectures.

\paragraph{N1 is the hardest stage.}
N1~F1 ranges from 0.1519 (YASA) to 0.5327 (BIOT) on EEG-only. It is the most variable stage and the primary beneficiary of better pretraining. N1 has low inter-rater agreement even among human experts, and its EEG signatures are subtle and brief.

\paragraph{The deployment cost of ECG over EEG is small, but the performance drop is large.} Figure~\ref{fig:deployment} summarizes this trade-off directly. Across the five models capable of ECG-only staging, switching from EEG to ECG costs 0.2798 macro~F1 for MOMENT, 0.3229 for SleepFM from-scratch, 0.3443 for SensorLM, 0.4032 for LaBraM, and 0.4151 for BIOT, averaging 0.3531. The direction of this drop is consistent across five architecturally distinct models with different pretraining regimes, even though its size varies by nearly 0.14 macro~F1 between the smallest and largest case. ECG\_ONLY needs only 33.3\% of EEG\_ONLY's raw data rate (Table~\ref{tab:datavol}), so this is the concrete trade-off a wearable-only deployment faces: a third of the data rate for roughly 41\% to 53\% of EEG-only's macro~F1, depending on model choice, visible directly in Figure~\ref{fig:scatter} as the leftward-and-downward shift from EEG-only to ECG-only across every model. We do not have power-consumption or wireless-bandwidth measurements for any specific device, so we report this as a reduction in raw channel data rate only, not as an estimate of battery life or transmission cost.

\FloatBarrier

\section{Limitations}\label{sec:limitations}

Our evaluation has several limitations. We use a fixed subsample of 350 subjects out of the more than 2,000 available in the full MESA Sleep cohort, chosen for a pilot-scale evaluation across six independently trained or fine-tuned models rather than as a hard data constraint; MESA itself does not limit us to this size. We have not tested whether the EEG-to-ECG accuracy gap holds, strengthens, or narrows on the full cohort, and a larger sample could also support demographic stratification that our current 350-subject split does not have room for. All results also come from a single dataset, MESA, so we cannot say whether the EEG-to-ECG accuracy gap we measure would hold across a different PSG cohort, age range, or recording setup. The held-out test split comprises 50 subjects, drawn from a single fixed 10-fold assignment with a fixed random seed, and is not stratified by demographic or clinical variables, such as apnea severity. A different split could shift individual model results by more than the differences we report between models. We report a single macro~F1 value per model per condition rather than a distribution across seeds or folds, so we cannot attach a confidence interval to the 0.2798-0.4151 range of EEG-to-ECG accuracy costs in Section~\ref{sec:discussion}, and cannot rule out that part of that spread reflects fine-tuning variance rather than a real architectural difference.

Our data-volume analysis (Table~\ref{tab:datavol}) reports only the raw signal size in bytes per second, derived from the channel count, sample rate, and ADC resolution. We do not have measured power consumption or wireless bandwidth figures for any wearable device, so we cannot translate this data rate reduction into an estimate of battery life or transmission cost. A raw data-rate comparison is a reasonable proxy for deployment burden, but it is not a substitute for a hardware measurement, and we treat it as such throughout this paper.

\FloatBarrier

\section{Conclusion}

We compared six sleep-staging models, spanning EEG-pretrained, general time-series, from-scratch, and classical strategies, under EEG, ECG, and EEG+ECG input, with no model architecture modified from its original form. EEG carries most of the sleep-staging signal: ECG alone yields near-zero F1 for N1 and N3, and adding ECG to EEG improves only one of six models. Across the five models capable of ECG-only staging, replacing EEG with ECG costs between 0.2798 and 0.4151 macro~F1, averaging 0.3531, while cutting the raw channel data rate to a third. This drop is consistent in direction across five architecturally different models, suggesting it reflects a real limit of what ECG can resolve about sleep stage rather than a property of any one architecture. For an edge-cloud deployment, where EEG requires clinic-grade electrodes that a wearable device cannot provide, this is the trade-off signal choice actually carries: substantially lower data rate against a substantial, consistent loss in accuracy.

%These results establish baselines for future work on sleep-based dementia biomarkers from wearable signals. They come from a 350-subject MESA subsample and a single held-out split, so they are a starting point rather than a final answer. The most direct next steps are to test this gap on a larger cohort and pair our raw data-rate comparison with actual power and bandwidth measurements from a wearable device.

\FloatBarrier

\section*{Acknowledgements}

This work was supported by the Horizon Europe AI4HOPE Project under Grant Agreement No. 101136769 and Shanghai Sci-tech Co-research Program Project under Grant (25HB2703300). All experimental work, data analysis, and research decisions were carried out by the authors. Claude (Anthropic) is used as an AI writing assistant to support editing and text structuring. All AI-assisted text was reviewed, revised, and verified by the authors before inclusion.

Compute resources were provided by CSC, Finnish IT Center for Science (Puhti supercomputer). MESA data were accessed through the National Sleep Research Resource (NSRR).

\bibliographystyle{IEEEtran}
\bibliography{references}

\end{document}